\documentclass[]{aa}
\usepackage{natbib}

\def\Msun{$M_{\odot}$}
\def\Rsun{$R_{\odot}$}
\def\SB9{$S\!_{B^9}$}
\def\kms{km~s$^{-1}$}

\bibpunct{(}{)}{;}{a}{}{,}

\usepackage[dvips]{graphicx}
\usepackage{times}
\newcommand\ignore[1]{} 
\usepackage{amsmath}

\begin{document}
\title{Spectroscopic binaries among Hipparcos M giants\thanks{Based on
observations carried out at the Swiss telescope installed at the
{\it Observatoire de Haute Provence} (OHP, France), and at the 1.93-m OHP
telescope}}
\subtitle{II. Binary frequency}
\titlerunning{Spectroscopic binaries among M giants}
\author{A.~Frankowski\thanks{Postdoctoral Researcher, F.N.R.S., Belgium.
Currently at Department of Physics, Technion-Israel Institute of
Technology, Haifa 32000, Israel}\fnmsep\inst{1}
\and B.~Famaey\thanks{Postdoctoral Researcher, F.N.R.S., Belgium}\fnmsep\inst{1}
\and S.~Van Eck\thanks{Research Associate, F.N.R.S., Belgium}\fnmsep\inst{1}
\and M.~Mayor\inst{2}
\and S.~Udry\inst{2}
\and A.~Jorissen\inst{1}
}
\institute{
Institut d'Astronomie et d'Astrophysique, Universit\'e
libre de Bruxelles, Facult\'e des Sciences, CP.~226, Boulevard du Triomphe, B-1050
Bruxelles, Belgium  
\and
Observatoire de Gen\`eve, Universit\'e de Gen\`eve, CH-1290 Sauverny, Switzerland
}

\date{Received date; accepted date}

\abstract
{This paper is the second in a series devoted to  studying the
properties of binaries with M giant primaries.}
{The binary frequency of field M giants is derived and
compared with the binary fraction of K giants.}
{Diagrams of the CORAVEL spectroscopic parameter $Sb$ (measuring the
average line width) vs. radial-velocity
standard deviation for our samples were used to define appropriate binarity
criteria. These then served
to extract the binarity fraction among the M giants. Comparison is made to
earlier data on K giant binarity frequency. The $Sb$ parameter is discussed in
relation to global stellar parameters, and the $Sb$ vs. stellar radius
relation is used to identify fast rotators.}
{We find that the spectroscopic binary detection rate among field M giants,
in a sample with few velocity measurements ($\sim2$), unbiased
toward earlier known binaries, is 6.3\%. This is less than half of the
analogous rate for field K giants. This difference
originates in the greater difficulty of finding 
binaries among M giants because of their smaller orbital velocity
amplitudes and larger intrinsic jitter and in the different distributions of K and M giants in
the eccentricity -- period diagram.
A higher detection rate was obtained in a smaller M giant sample with more
radial velocity measurements per object: 11.1\%  confirmed plus 2.7\% possible
binaries.
The CORAVEL spectroscopic parameter $Sb$ was found to correlate better with
the stellar radius than with either luminosity or effective temperature
separately. Two outliers of the $Sb$ vs. stellar radius relation, HD~190658
and HD~219654, have been recognised as fast rotators. The rotation is
companion-induced, as both objects turn out to be spectroscopic binaries.}
{}
\maketitle

\section{Introduction}
\label{Sect:Intro}

This paper is the second in our series discussing the spectroscopic-binary
content of a sample of M giants drawn from the Hipparcos Catalogue (ESA 1997),
for which CORAVEL radial velocities have been obtained in a systematic way
\citep{Udry-1997,Famaey-2005}. The main driver behind such an extensive
monitoring effort lies of course with the kinematics; indeed, the
kinematical properties of the present sample of M giants have been fully
analysed by \citet{Famaey-2005}. But this large amout of material may also be used
to search for binaries, which is discussed in a series of
three papers. In Paper~I \citep{Famaey-2008:b}, we present the
radial-velocity data and orbital
elements of newly-discovered spectroscopic binaries, and also discuss the
intrinsic variations sometimes causing pseudo-orbital variations. As a
follow-up, the present paper presents the first attempt to derive the
frequency of spectroscopic binaries in such an extensive sample of M giants.
A side topic of this paper is the identification of fast rotators among M
giants, using the CORAVEL line-width parameter $Sb$. This way of finding
binaries is independent of radial velocities. Finally, we investigate the
relation between the CORAVEL line-width parameter $Sb$ and stellar parameters
and conclude that $Sb$ is best correlated with the stellar radius.
In Paper~III \citep{Jorissen-2008}, more evolutionary considerations are
presented, based on the analysis of the eccentricity -- period diagram
of our sample of M giants.

\section{Binary frequency among M giants}
\label{Sect:frequency}

The sample used for the present study and its three subsamples are
extensively described in Paper~I. Sample I consists of 771
stars from the Hipparcos survey, for which at least 2 radial-velocity
measurements have been obtained \citep{Udry-1997}. Sample II consists of
about one-third sample I stars (254 stars), for which at least 4
measurements have been obtained. Sample III consists of 35 sample II stars
with $\sigma(V_r) > 1$~\kms, and sample IV of 138 sample II stars with
$\sigma(V_r) < 1$~\kms.

\subsection{Spectroscopic binaries (sample I)}
\label{Sect:freqI}

\begin{figure*}[]
  \includegraphics[width=\columnwidth]{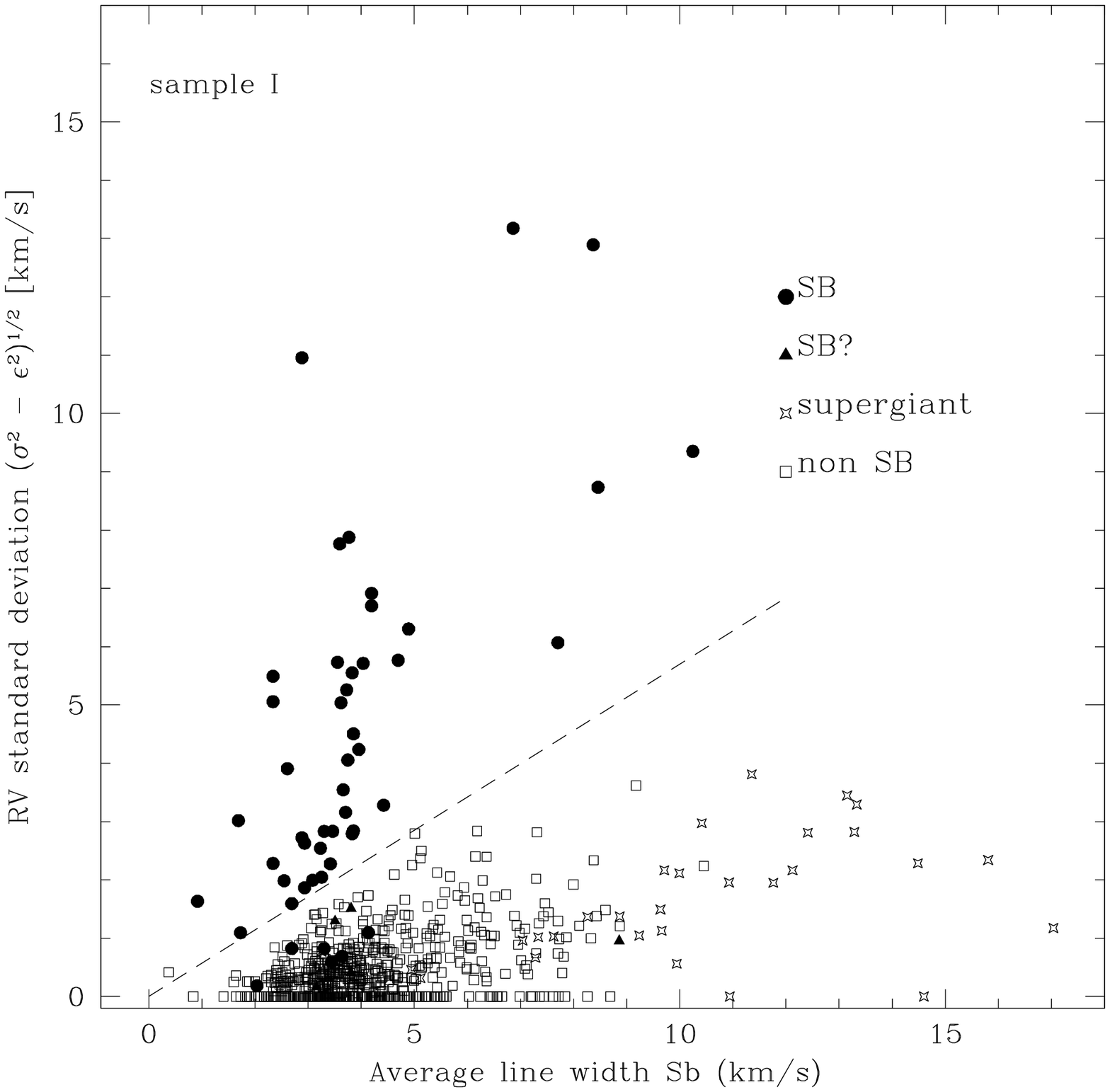}
  \includegraphics[width=\columnwidth]{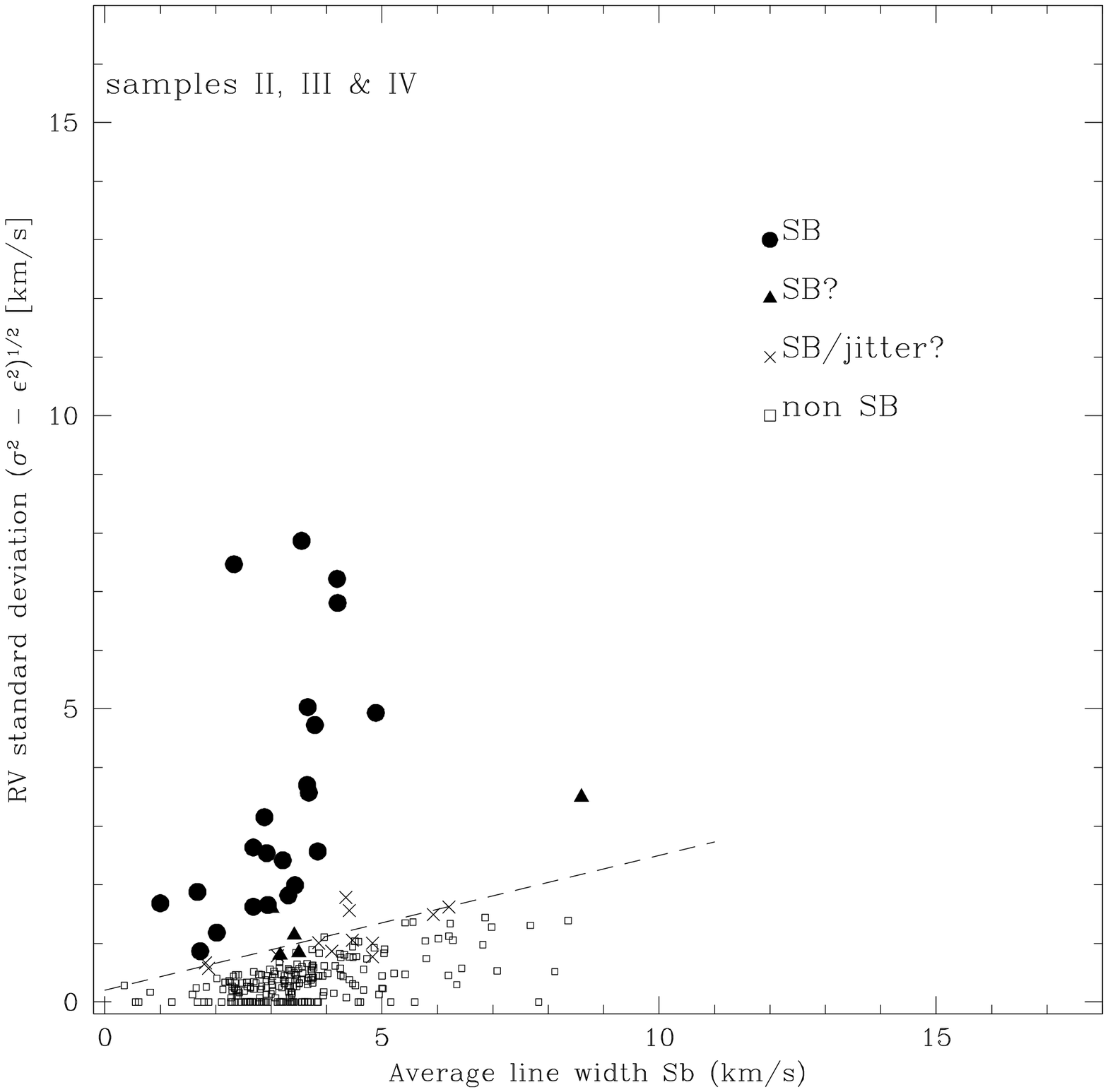}
  \includegraphics[width=\columnwidth]{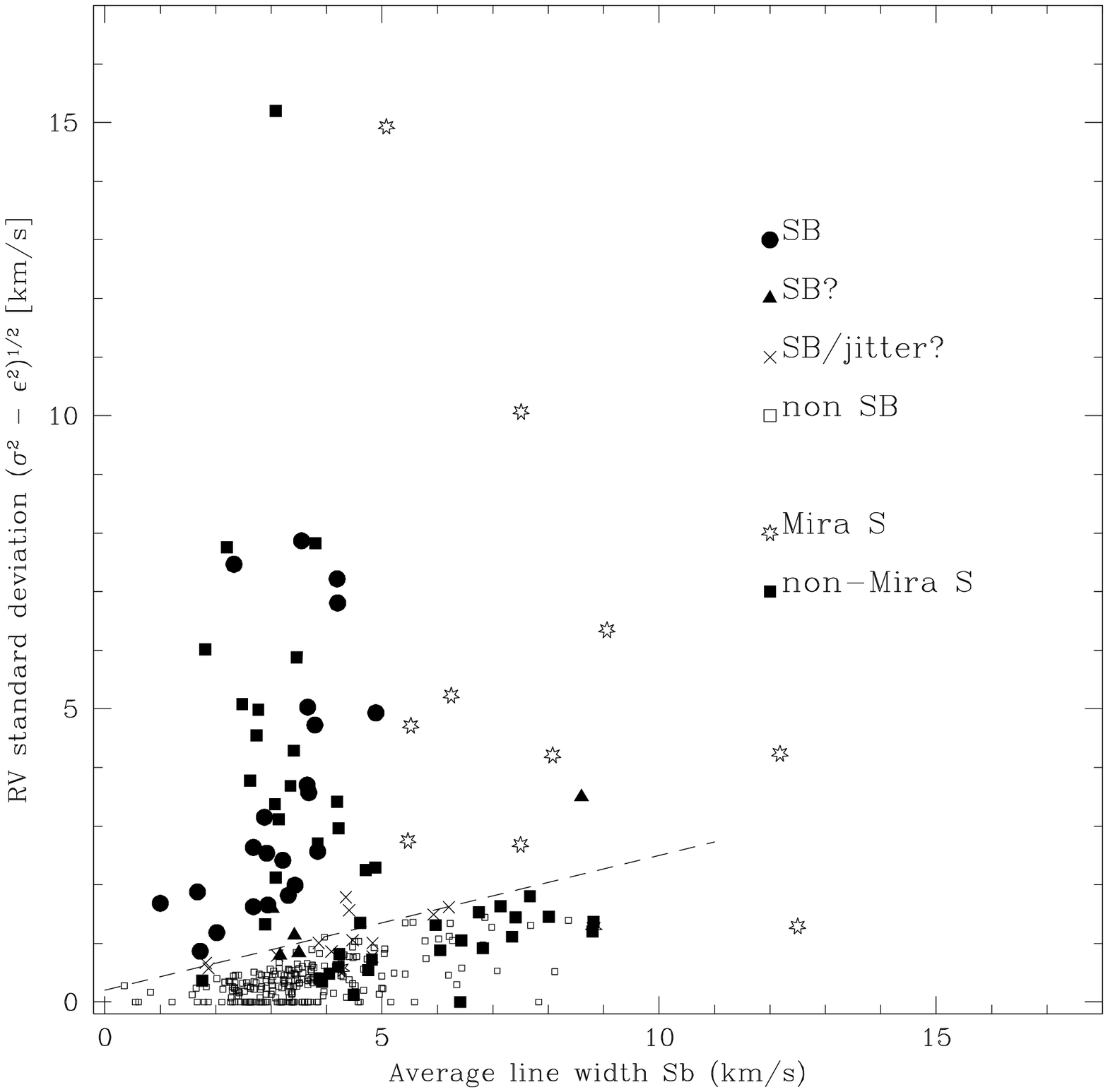}
  \includegraphics[width=\columnwidth]{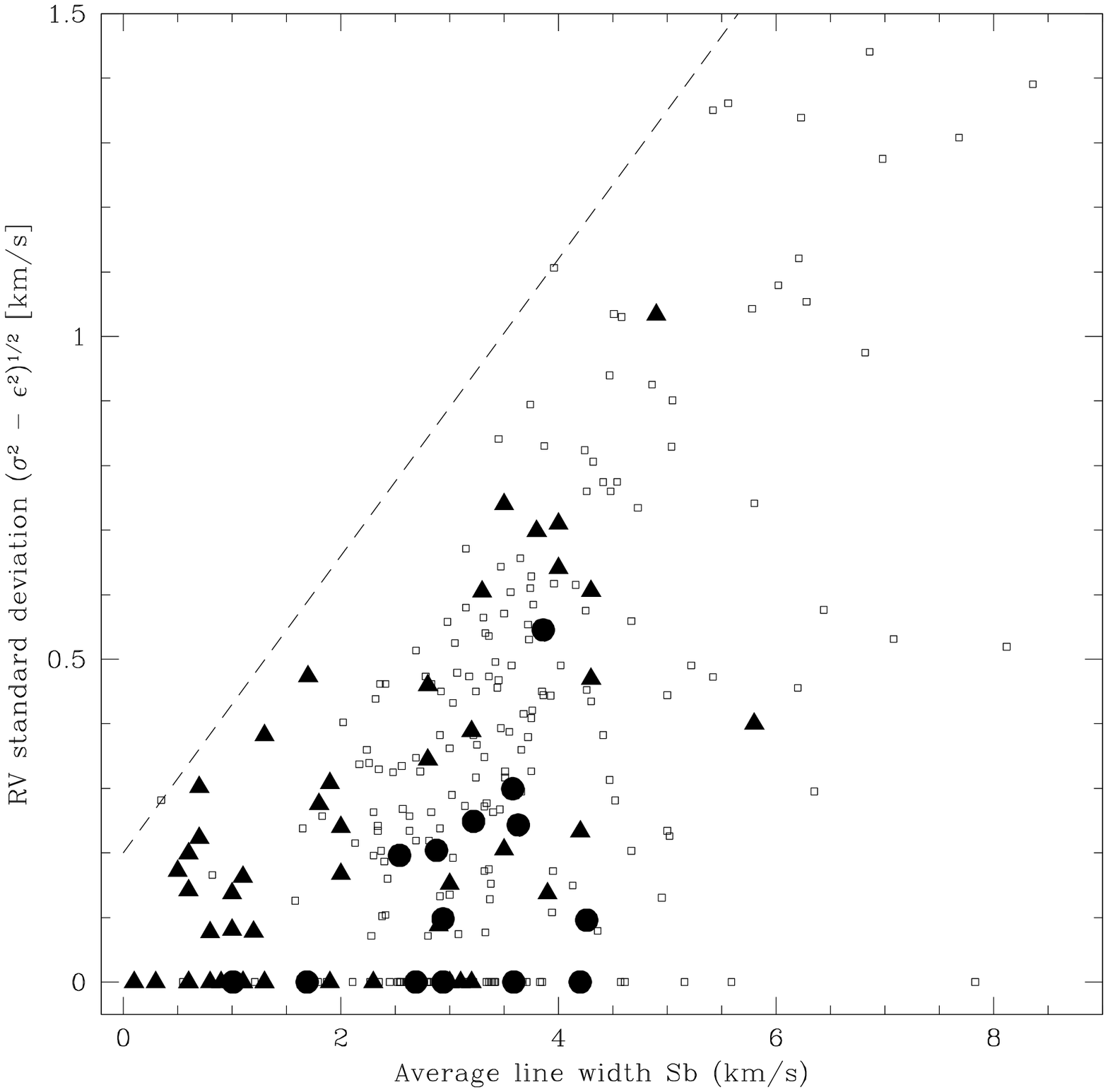}
  \caption{\label{Fig:Sb-sigma} 
(a) Top left panel: 
The radial-velocity
standard deviation $\sigma_0(Vr)$ (corrected for the average instrumental
error $<\!\! \epsilon \!\!>$) as a function of the average intrinsic line width
$Sb$, for sample~I. For non-binary stars (open squares), the maximum
radial-velocity jitter increases with $Sb$.  Filled circles denote binary
stars, as identified from their position above the dashed line, combined
with the more detailed information gathered from samples II, III, and IV.
Diamonds refer to supergiants.
(b) Top right panel: Same as top left panel but for samples II, III and IV.
Symbols are as follows: Filled circles: SBs; crosses: SB or jitter;
diamond: jitter.
The dashed line very clearly marks the limit between SBs (above)
and stars with intrinsic radial-velocity jitter (below).  
(c) Bottom left panel: Same as (b), adding
extrinsic
S stars from
\citet{Udry-98a}. Filled triangles: non-Mira S stars;
star symbol: Mira S stars.
(d) Bottom right panel: same as (c) after removing the spectroscopic
binaries with no orbit
and replacing the radial-velocity standard deviation
for orbital solutions by the standard deviation of their O-C residuals,
corrected for the average instrumental error (filled circles).
Filled triangles correspond to orbits for barium and extrinsic S stars from
\citet{Udry-98a,Udry-98b}.
The dashed line and open squares are the same in panels (b), (c), and (d).}
\end{figure*}

As explained in Paper~I, the detection of spectroscopic binaries (SBs) among
M giants cannot rely solely on a $\chi^2$ test comparing the radial-velocity
standard deviation $\sigma(Vr)$ to the average instrumental error
$<\!\epsilon\!>$, because of the mass motion existing in the atmospheres of
these stars (either due to convection or to pulsation), causing intrinsic
radial-velocity variations. In Paper~I, it was indicated that the parameter
$Sb$, measuring the average width of spectral lines (corrected for the
instrumental width), provides a useful tool to distinguish intrinsic
radial-velocity jitter from  orbital motion (see 
Fig.~\ref{Fig:Sb-sigma})\footnote{Note that the present figure differs slightly  from the one 
published by
\citet{Famaey-2005}, which is in error.}.
This is because the $Sb$ (i.e., line-width) parameter is a good tracer of
the evolutionary state of giant stars (as is shown in
Sect.~\ref{Sect:Sb}), and extensive mass motions are occurring in the
envelopes of highly evolved giants. These mass motions cause the radial
velocity jitter, which is thus large in highly evolved giants, where $Sb$ is
the largest. 

In the $Sb-\sigma_0$ diagram, where
$\sigma_0(Vr) = (\sigma^2 - <\! \epsilon \!>^2)^{1/2}$,
stars are flagged as binaries if they fall above a `dividing line'
(Fig.~\ref{Fig:Sb-sigma}). The binaries identified from
the literature and from
the more intensive monitoring of subsamples II, III, IV (Sect.~3.1 and Table~4 of Paper~I)
have been used to define that boundary: the vast majority of these binaries
fall above that boundary, while at the same time, confirmed non-binary
stars are not found in that region.
A dividing line with a slope of 0.57 fulfills these conditions.
There is one apparent exception in the lower left corner, HD 40282, which
in follow-up observations turned out not to exhibit orbital motion.
It may be a statistical fluctuation in the measured jitter. (Compare to
the discussion below, where a Gaussian distribution for the radial velocity
jitter is assumed.)
Besides, its status as an exception depends on the exact definitions
of $Sb$ and of the dividing line.
It must be stressed that this slope for sample~I is steeper than that
obtained for the
dividing line in diagrams where the radial-velocity standard deviation is
computed from more extended data sets characterising samples II, III,
and IV (see the corresponding panels in Fig.~\ref{Fig:Sb-sigma}).

As these lines are defined by the observed distribution of the
{\it single-star} objects,
the difference in the position of the dividing lines can be explained
by a combination of two factors: the number of radial-velocity data points
for a given object and the total number of objects in a sample.

First, every $\sigma_0$ value is an observational
estimate of the true standard deviation $\tilde{\sigma}$.
It is intuitive that the spread in the values of such an estimator is
broader when the number of datapoints (here: radial-velocity measurements)
is smaller. 
Let us assume that the radial-velocity jitter follows a Gaussian distribution,
with variance $\tilde{\sigma}^2(Sb)$ at any given $Sb$ value.
An estimator of $\tilde{\sigma}^2$, the sample variance ${s^2}_{N-1}$
($= {\sigma_0}^2$), for an underlying normal distribution
follows a scaled $\chi^2$ distribution with $N-1$ degrees of freedom,
where $N$ is the number of measurements used to compute ${s^2}_{N-1}$.
The mean of this $\chi^2$
distribution is $\tilde{\sigma}^2$
and its variance $2\tilde{\sigma}^4/(N-1)$.
The sample standard deviation, ${s}_{N-1}$ ($= \sigma_0$), which is an
estimator of $\tilde{\sigma}$, in turn follows a scaled $\chi$ distribution with
$N-1$ degrees of freedom, an expected value of $c_4 \tilde{\sigma}$, and a variance
of $(1-{c_4}^2)\tilde{\sigma}^2$.
Here, $c_4 = \sqrt{2/(N-1)} \Gamma(N/2)/\Gamma((N-1)/2)$ and $\Gamma(x)$ is
the gamma function.
Therefore the standard deviation of the variable $s_{N-1}$ for $N=2$
(as in sample~I) is $\approx1.55$
times more than for $N=4$ (a value prevailing for samples II, III, and IV).
However, the
effect of this for the upward spread of $s_{N-1}$ is partially offset by
the expected value of ${s}_{N-1}$ being lower for $N=2$
compared to $N=4$, by a factor of $\sqrt{3}/2$.

The other contributing factor is that sample~I contains about 3 times more
objects than sample II. The numbers of (apparently) single objects
differ similarly (728 vs. 227 using results from this section --see below-- and from
Sect.~\ref{Sect:freqII}),
which makes the low-probability,
large-deviation tail of the $\sigma_0$ distribution more populated in sample~I.
At some value ${s}_{N-1, \mathrm{crit}}$, the probability of finding
 a value of
${s}_{N-1}>{s}_{N-1, \mathrm{crit}}$ for any given object becomes less than $1 /$(sample size).
Finding such high ${s}_{N-1}$ is intuitively `unlikely' - in fact,
for large samples this $1 /$(sample size) limit is equivalent to
a probability of $1/$e $\approx 0.37$ that no value of ${s}_{N-1}$
{\em in the whole sample} is found above ${s}_{N-1, \mathrm{crit}}$.
For $N=2$, this critical value, ${s}_{N-1, \mathrm{crit}}$,
above which the probability Prob$({s}_{N-1})$ becomes less than $1 / 728$
is 3.20. For Prob$({s}_{N-1}) < 1/227$,
${s}_{N-1, \mathrm{crit}}$ is 2.85. The ratio of these values is 1.12,
which translates into a ratio of the ranges of $s_{N-1}$ easily populated
by the two samples
(purely due to their size, as $N=2$ was assumed for both).
In fact, these two effects combine: an $N=4$ case with 227 objects
(sample II) has ${s}_{N-1, \mathrm{crit}} = 2.09$, and comparing the ranges
of $s_{N-1}$ populated by these samples gives a ratio of
$3.20 / 2.09 = 1.53$.
Applying this factor to the dividing line in the $Sb-\sigma_0$ diagram of
sample~I, of slope 0.57, gives an expected slope of 0.372 for sample~II.
This should be compared to an
observed value of $~0.3$--0.4, which can be derived for samples II,
III, IV when a dividing line going through point (0,0) is adopted, just as
for sample~I (Fig.~\ref{Fig:Sb-sigma}a).
The final detailed position of samples' II, III, IV dividing
line is defined in a slightly different way than for sample I, namely from an
$Sb-\sigma_0$ diagram with the orbital motions of known binaries subtracted
(Fig.~\ref{Fig:Sb-sigma}d). This adopted line has a positive intercept,
resulting in less of a slope: $0.23 Sb + 0.2$.
Nevertheless, the agreement between the above reasoning and the relative
positions of the observational dividing lines is evident.

In summary, stars falling {\em above} the dividing line may be safely
flagged as binaries as long as $Sb < 5$~\kms\ (i.e. for non-Mira stars).
For higher $Sb$ values, the dividing line loses its diagnosis power,
since $Sb > 5$~\kms\ is the realm of Mira and semi-regular variables which
may have intrinsic radial-velocity variations with standard deviations of
the order of 10~\kms\ (Fig.~\ref{Fig:Sb-sigma}c). 
For those stars, it is almost impossible to use radial
velocities  to detect binaries. Methods based on the proper motion may be
used instead (see Sect.~\ref{Sect:Deltamu}).

Using the criterion based on the dividing line, 43 spectroscopic binaries
were identified in sample~I (upper left panel of
Fig.~\ref{Fig:Sb-sigma}), corresponding to a binary frequency of
43 / 771 = 5.6\%. The list of binary stars in sample~I may be found in
Table A.1\footnote{The star HIP 26247 = RR Cam = BD +72$^\circ$275
was erroneously flagged as a binary in \citet{Famaey-2005}, as the result of a
confusion between the CORAVEL coding for BD +72$^\circ$275, and for the binary
star J275 in the Hyades} of \citet{Famaey-2005}.

Because the detection of binaries is difficult among Miras (see
Sect.~\ref{Sect:freqII}), we also compute
the binary frequency for non-Miras (by limiting sample~I to
$Sb < 5$~\kms), namely
38 / 603 = 6.3\%. That value will be of interest for comparing with
K giants as discussed in Sect.~\ref{Sect:K} and Table~\ref{Tab:frequency}.

\subsection{Proper-motion binaries}
\label{Sect:Deltamu}

\begin{table}
\caption[]{\label{Tab:deltamubinaries}
Supplementary `proper-motion binaries' found in sample~I using the 
method described in \protect\citet{Frankowski-2007} and extracted 
from their Table~2.  
}
\begin{tabular}{llllllllll}
\hline
\hline\\
HD/BD & HIP & Rem. & \\ 
      &              \\ 
\hline\\
5006 & 4100 & SB$^a$ \\
+58 501  & 12416 & supergiant \\
51802 & 37391 & \\
52554 & 33929 & \\
96572 & 54571 & SB$^a$ \\
103945 & 58377 \\
108907 & 60998 & SB$^a$ ($P = 1714$~d)\\
130218 & 72268 \\
162159 & 87114 \\
190658 & 98954 & SB$^a$ ($P = 198.7$~d)\\
       &       & + visual companion at 2.3'' \\
+39 4208 & 101023 & supergiant\\
199871 & 103528 & SB$^a$ ($P = 840$~d)\\
209026 & 108588 \\
219981 & 115191 \\
220088 & 115271 & SB$^a$ ($P = 2518$~d)\\\\
\hline\\  
\end{tabular}

$^a$ the star is also
flagged as spectroscopic binary from the radial-velocity data.
 \end{table}

A few more
binary stars have been found from the comparison of
the short-term Hipparcos proper motion with the long-term Tycho-2
proper motion, since only the former is altered by the orbital motion
\citep[for orbital periods in the range 1000 - 30\ts
   000~d;][]{Frankowski-2007}. These `proper-motion-difference binaries' in
 sample~I are
 listed in Table~\ref{Tab:deltamubinaries}. They are not added, however, to the {\it spectroscopic} binary frequency derived in this paper, because the detection efficiency of these binaries involves different selection biases than those associated with the detection of SBs. For instance, proper-motion binaries are efficiently detected only among stars with parallaxes in excess of $\sim 20$ mas, which are rare among M giants. 

\subsection{Fast rotators identified from the CORAVEL line-width parameter $Sb$}
\label{Sect:Sb}

Another method of finding binaries among M giants involves
identifyingfast rotators. Single M giants are not expected to rotate
fast \citep[see e.g.][]{DeMedeiros-1996,DeMedeiros-1999}, so that fast
rotators can be ascribed to spin-up processes operating in tidally
interacting systems \citep[like RS~CVn among K giants; ][]{DeMedeiros-2002}.
This identification is possible by using the CORAVEL parameter $Sb$
(measuring the average width of spectral lines
corrected for the instrumental width) to search for outliers in the
relation between this parameter and the stellar radius, $R$, as
shown in Fig.~\ref{Fig:R_Sb}. In previous papers, $Sb$ was used as
a luminosity proxy \citep[][ where it was used
to distinguish extrinsic (Tc-poor) from intrinsic (Tc-rich) S
stars]{Jorissen-VE-98,VanEck-Jorissen-00}. Now we have found that the
tightest correlation is in fact obtained between $Sb$ and the stellar
radius, rather than between $Sb$ and luminosity, $L$,
or effective temperature, $T_{\rm eff}$.
It is the $Sb$ -- radius relationship that is therefore the most
appropriate for looking for outliers corresponding to fast rotators. 

Before embarking on the search for fast rotators, it is useful to first test
the relationship between $Sb$ and stellar luminosity, since
maximum-likelihood distances are available for all the stars of sample~I
\citep{Famaey-2005}. A possible relation to temperature may also be
checked using the $(V-I_C) - T_{\rm eff}$ calibration of \citet{Bessell-98}
with $(V-I_C)$ from \citet{Platais-2003}.

\begin{figure}
\includegraphics[width=\columnwidth]{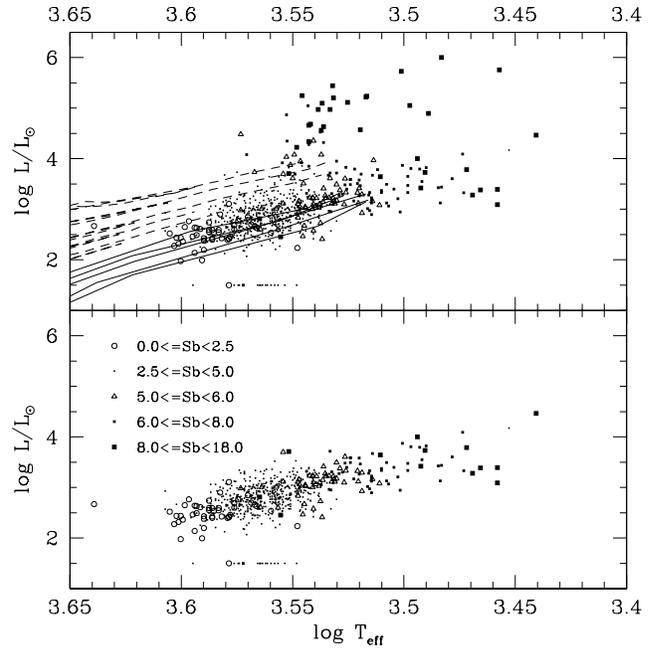}
\caption[]{\label{Fig:HR_Sb}
Upper panel: The HR diagram for the full sample~I \citep[but the visual
binaries, identified by flag 4 in Table~A.1 of][ see text for an explanation
about how luminosity and $T_{\rm eff}$ were derived]{Famaey-2005}.
The different symbols denote $Sb$ ranges as indicated by the labels. Lower panel: Same as upper panel after eliminating stars
flagged as `Y' (young, supergiant stars) by \citet{Famaey-2005}. Since
maximum-likelihood distances, and hence luminosities, could not be derived
by \citet{Famaey-2005} for spectroscopic binaries with no orbit, these
binaries are displayed in both panels at an arbitrary luminosity of
$\log L/L_{\odot} = 1.5$. Evolutionary tracks from \citet{Charbonnel-1996}
(up to the tip of the RGB; no overshooting) for masses 0.9, 1.0, 1.25, 1.5,
and 1.7 M$_\odot$ are shown as solid lines. Dashed lines are tracks from
\citet{Schaller-1992} for masses of 2.0, 2.5, 3.0, 4.0, and 5.0 M$_\odot$.
For low-mass stars, the tip of the RGB corresponds to $Sb = 5$ to 6~\kms.
A colour version of this plot is available in the electronic version only.}
\end{figure}

\begin{figure}
\includegraphics[width=\columnwidth]{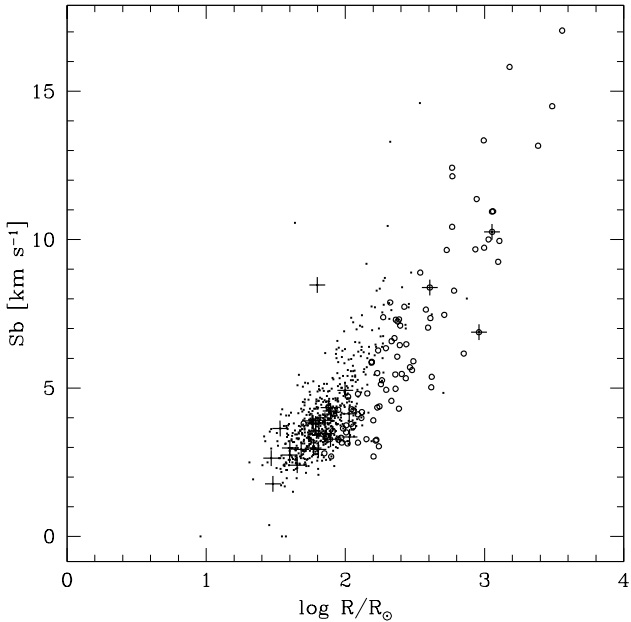}
\caption[]{\label{Fig:R_Sb}
The relationship between $Sb$ and the stellar radius, derived from the
Stefan-Boltzmann law and the HR diagram of Fig.~\ref{Fig:HR_Sb}.
Open circles denote stars flagged as `Y' (young) by \citet{Famaey-2005} and
large crosses identify binaries. The outliers with large $Sb$ in this diagram
are supposedly binaries, because tidal interactions in the binary spun up or
slowed down the giant's rotation. Binaries with no orbits are not included in
this figure, since they have no maximum-likelihood distances available either.
}
\end{figure}

Figure~\ref{Fig:HR_Sb} presents the Hertzsprung-Russell diagram for sample~I
\citep[excluding visual binaries, identified by flag 4 in Table~A.1
of][]{Famaey-2005}, where  $T_{\rm eff}$ has been derived from the
above-mentioned calibration, and the luminosity derives from
$M_{\rm bol} = M_V - (V-I_C) + BC_{I_C}$, with the bolometric correction
$BC_{I_C}$ from \citet{Bessell-98} and $M_V$ from Tycho-2 $V_{T2}$, combined
with the maximum-likelihood distance from \citet{Famaey-2005}. Two sequences
are readily apparent: the lower, most populated one involves giant stars,
and the upper one, containing only stars with high $Sb$ values, involves
supergiant stars. All stars flagged as `Y' (young) by \citet{Famaey-2005}
belong to the upper sequence. 
The comparison of the evolutionary tracks with the location of the M giants
in the HR diagram of Fig.~\ref{Fig:HR_Sb} reveals that they have typical
masses in the range 1.2 -- 1.7~\Msun.   
As far as $Sb$ is concerned, it increases monotonously along the sequences; 
however, the existence of the two sequences makes it possible to check and
exclude the correlation of $Sb$ with either $T_{\rm eff}$ or luminosity
alone. 
Instead, $Sb$ correlates well with the stellar radius (Fig.~\ref{Fig:R_Sb}),
which is a combination of $T_{\rm eff}$ and $L$.

There are two stars (HIP 8175 = HD 10696 and
HIP 98954 = HD 190658)\footnote{
HD~10696 is not displayed on the $R - Sb$ plot of Fig.~\ref{Fig:R_Sb}, because that star has only one radial-velocity measurement, and was thus removed from sample~I.} with moderate radii ($\log R/R_{\odot} = 1.6$ -- 1.8)
and with $Sb$ much larger than expected at the corresponding radii. At larger
radii, $\log R/R_{\odot} > 2.2$, the relation between $R$ and $Sb$ becomes
scattered -- basically because the CORAVEL cross-correlation dip from which
$Sb$ is derived becomes shallow and asymmetric -- and the identification of
outliers is no longer possible.  

HD 219654 is another example of star with too
large an $Sb$,  found  in Fig.~\ref{Fig:HR_Sb} in the form of a larger square
amidst small ones at $\log L/L_{\odot} = 1.5$. 
It does not appear in Fig.~\ref{Fig:R_Sb}, because
\citet{Famaey-2005} did not derive maximum-likelihood distances for that
star; therefore, its luminosity and hence radius cannot be derived.
The reason for the lack of distance is that the maximum-likelihood
distance estimator defined by \citet{Famaey-2005} makes use of the
radial velocity, and no reliable centre-of-mass velocity could be defined
for those stars that appeared to be binaries with rather large velocity
amplitude and no orbit available
\citep[objects with flag `0' in][]{Famaey-2005}.

All fast-rotator candidates are listed in Table~\ref{Tab:Sb}.
Of these, HD 190658 and HD 219654 appear to be genuine fast rotators. The calibration of $Sb$ in
terms of the rotational velocity $V \sin i$ is from \citet{Benz-1981}.
Fast rotation is expected in binaries close enough for tidal interactions to
spin up the giant's rotation \citep[for K giants, the critical period below
which synchronisation takes place is 250~d; see][]{DeMedeiros-2002}.
One clear example of this behaviour is provided by HD~190658, the M-giant
binary with the second shortest period known (199~d; Table~8 of Paper~I),
which exhibits ellipsoidal variations as well \citep{Samus-1997}.
Its $Sb$ of 8.5~\kms\
formally
corresponds to  a rotational velocity of
$V \sin i = 15$~\kms.
Correcting for a typical (supposedly turbulent) broadening of 3.5~\kms\
for giants of this radius (estimated from Fig.~\ref{Fig:R_Sb}) yields
$V \sin i = 10$~\kms.
The radius derived from assuming synchronisation is then
$R \sin i = 39$~\Rsun.
With $\log T_{\rm eff} = 3.565$ and $\log (L/L_{\odot}) = 2.8$, the radius
deduced from Stefan-Boltzmann law is 62.2~\Rsun, which implies an orbit
inclined at $i = 39^\circ$. Rotation slower than synchronous would
entail an inclination closer to edge-on.

\begin{table*}
\caption[]{\label{Tab:Sb}
Binaries inferred from the high rotational velocity of the giant primary. 
}
\begin{tabular}{rrllllllllp{4cm}}
\hline
\hline
HD  & HIP & SB?$^a$  & $\sigma_0$ & $Sb$ &  $V \sin i$ & $V-I$  & $T_{\rm eff}$  & $M_V$  & $R$  & Rem\\
    &     &      & (\kms) & (\kms) & (\kms) & & (K) & & ($R_{\odot}$) \\
\hline
10696 & 8175 & - & - & 10.5 &  - & 2.3  & 3590 & 0.51 & 43 & not a fast rotator; $Sb$ contaminated by close (0.16'') visual companion\\
190658& 98954 & y & 8.7 & 8.5 &   10 & 2.1 & 3670 & $-0.67$ & 63   & Ellipsoidal variable, spectroscopic binary with $P = 199$ d\\
219654 & 114985 & y & 6.1 & 7.7 & 14 & 1.9 & 3730 & - & -   & not listed in Table~4b of Paper~I because it only belongs to sample~I\\
\hline
\end{tabular}

$^a$ This column provides the diagnostics of spectroscopic binarity
based on the location of the star in the $Sb - \sigma_0(Vr)$ diagram
(Fig.~\ref{Fig:Sb-sigma}).\\
HD 219654 : SB: JD 2\ts448\ts843.634:     $5.22\pm 0.4$~\kms,  JD 2\ts449\ts676.270:   $13.84\pm 0.3$~\kms

\end{table*}

The radial-velocity data clearly reveals that HD 219654 is a spectroscopic
binary (see its location on the $Sb - \sigma_0(Vr)$ diagram in
Fig.~\ref{Fig:Sb-sigma}a from the
parameters listed in Table~\ref{Tab:Sb}). An outlying $Sb$ value thus
appears to be a good way to detect a close binary (with periods up to
$\sim 300$~d).
However, there is an important caveat: spurious detections are possible when
a close visual companion contaminates
the giant spectrum and makes it composite with seemingly broadened lines.
HD~10696 (=~HIP~8175) is a such case, since it is a close visual binary
detected by Hipparcos ($Hp = 8.54$ and 9.09 with a separation of 0.16
arcsec). It is  therefore likely that in this case the large $Sb$ comes
from the composite nature of the spectrum rather than from rapid rotation
of the M giant star.

\subsection{Spectroscopic binaries (samples II, III, IV)}
\label{Sect:freqII}

The binary frequency derived for sample~II, 
considering the supplementary knowledge gained from the extensive
monitoring of the set of stars that had $\sigma(Vr) >
1$~\kms\ (corresponding to sample III), 
and from the $\sigma_0(Vr)$ -- $\sigma(Hp)$ diagram (where $\sigma(Hp)$ 
is the standard deviation of the Hipparcos $Hp$ magnitude; see Paper~I),
amounts to 17/254 {\it certain} binaries, or 6.7\%, 
with a  binomial error of 1.6\%. On top of these, 3 stars 
are possibly binaries, or 1.2\%. To detect very long-period binaries among
sample~II, 138 of the 219 stars not present in sample~III (spanning the
right-ascension range 0$^h$ - 16.5$^h$ and not already flagged as SBs)
have been re-observed, long after the last measurement of sample~II
(to make up sample~IV).
With this additional measurement, 5 binaries (and 2 suspected binaries)
have been found (Table~2 of Paper~I), adding a fraction of
$\frac{5\times(219/138)}{254}$=3.1\% to the total frequency.
Therefore, the estimated fraction of spectroscopic binaries
among M giants is 6.7+3.1 = 9.8\%, plus 2.4\% of suspected binaries.  
This frequency is 1.6 times higher than the value obtained for sample~I, and this is
related to the difference in binarity detection limits between samples
(Sect.~\ref{Sect:freqI}).

The above frequency should still be corrected for the selection bias acting
against detecting binaries with a Mira component.
Exact accounting for binaries among Mira variables is difficult.
We argue that it is almost impossible to find spectroscopic binaries
among Mira variables \citep[and indeed very few are known;][]{Jorissen-03}. 
From the period-luminosity-radius relationship for Miras 
\citep{Wood-1990}, it is
possible to evaluate the minimum admissible orbital period for a Mira
of a given pulsation period in a detached system (assuming fundamental
pulsation and masses of 1.0 and 0.75~\Msun\ for the
 Mira star and
its companion, respectively; the conclusions are not very sensitive to
these masses, fortunately): it ranges from about 1000~d for a Mira
pulsating with 200~d period to about 2200~d for a 600~d
pulsator \citep[see Fig.~9.4 of ][]{Jorissen-03}. These orbital periods
translate into radial-velocity semi-amplitudes smaller than about
10~\kms. Given that some of these stars have intrinsic
radial-velocity jitters of the order of several~\kms\
(Fig.~1 and Alvarez et al. 2001)\nocite{Alvarez-2001}, 
the difficulty of detecting those
spectroscopic binaries using radial velocity techniques
immediately becomes clear. An alternative, astrometric method based on the comparison
between long-term Tycho-2 proper motions (not altered by possible
orbital motions) and short-term Hipparcos proper motions
\citep[][ and Sect.~\ref{Sect:Deltamu}]{Wielen-1999} has not
yielded any positive detection \citep[because most Mira stars are
quite distant, and thus have
parallaxes that are too small for this method to be applied 
meaningfully;][]{Frankowski-2007}. 

In the $Sb - \sigma_0(Vr)$ diagram, Mira variables have $Sb > 5$~\kms, and
Fig.~\ref{Fig:Sb-sigma} reveals that stars with $Sb \ge 5$~\kms\ represent a
minor fraction of the total sample (less than 10\%, as shown by the solid
line in Fig.~\ref{Fig:binfreq_KS}). Therefore, any correction factor will have a limited impact, fortunately.
A simple way to estimate the binary frequency among M giants,
circumventing  
the difficulty of detecting spectroscopic binaries involving Mira
variables, is to restrict the sample to the 225 stars having $Sb < 5$~\kms,
for which the efficiency of binary detection is good. The binary frequency
then becomes $0.098 \times 254/225 = 0.111$, to which a fraction of 0.027
of suspected binaries should be added.

Finally, our samples contain a few symbiotic and suspected
symbiotic stars: 5 in sample~I (EG And, T CrB, CH Cyg, V934 Her, and 4 Dra)
and 2 in sample~II  (CH Cyg and 4 Dra). Their hot companions are
most likely accreting white dwarfs, so their inclusion in the binary
statistics of M giants is disputable, as in this case they correspond to the
{\em second} M giant stage in the system's evolution. Nevertheless, EG And was flagged as
binary by the sample~I binarity criterion. If the symbiotic systems are excluded in the 
evaluation of the binary frequency for sample~II, this frequency would decrease slightly to 0.106.
(Only 4~Dra has to be excluded, since CH Cyg, with $Sb=7$~\kms, above the 5~\kms\
threshold, is already excluded from the above estimate.) Readers
  involved in binary population synthesis should thus note that the
  binary frequency of 0.111 obtained above corresponds to systems with
  an M giant primary and a main sequence {\it or} white dwarf
  companion. There is no guarantee, however, that the frequency of
  0.106 (obtained above by excluding symbiotic systems) does
  correspond to the  binary frequency for systems with main-sequence
  companions alone. It cannot be excluded that M giants with white
  dwarf companions hide among non-symbiotic systems, especially for M
  giants at the base of the red giant branch (RGB) where mass loss is weak. 
This issue will be developed further in Sect.~3 of Paper~III.

\subsection{Binaries above $Sb = 5$~\kms}
\label{Sect:above5}

A striking feature of Fig.~\ref{Fig:Sb-sigma} is the nearly complete absence
of stars above the dashed line when $Sb \ge 5$~\kms\ in sample~II.
Sample~I also exhibits an (apparent?) lack of stars above the dashed line
beyond $Sb = 5$~\kms.
In fact, 4 of the 5 binaries observed above
$Sb = 5$~\kms\ in sample~I (Fig.~\ref{Fig:Sb-sigma}) are
VV-Cep-like binaries  \citep{Cowley-1969}, flagged as belonging to the young
(`Y') group by \citet{Famaey-2005}:
\begin{itemize}
\item[-] HD 42474 (WY Gem; $Sb = 8.4, \sigma_0 = 12.9$~\kms): a VV Cep type
binary, consisting of a M2 supergiant and an early B type
main-sequence star \citep{Leedjaerv-1998};\\
\item[-] HD 190658 (V1472 Aql; $Sb = 8.5, \sigma_0 = 8.7$~\kms), an 
ellipsoidal or eclipsing binary of type M2.5 \citep{Lucke-Mayor-1982};\\
\item[-] HD 208816 (VV Cep; $Sb = 10.2, \sigma_0 = 9.3$~\kms);\\
\item[-] BD +61$^\circ$08 (KN Cas; $Sb = 6.9, \sigma_0 = 13.2$~\kms): a VV-Cep-type
binary \citep{Cowley-1969}.
\end{itemize}
The last one, HD~219654 (V349 Peg; $Sb = 7.7,
\sigma_0 = 6.1$~\kms), is a giant rather than a supergiant
star. It is nevertheless located among supergiants because it is a fast rotator, spun up 
by tidal interactions in a close binary, as discussed in Sect.~\ref{Sect:Sb}.

\begin{table*}
\caption[]{\label{Tab:hyper}
Results of a hypergeometric test checking the $H_0$ hypothesis that the
frequency $p$ of stars above the dividing line is the same on both sides
of the $Sb$ threshold value. 
}
\begin{tabular}{ll|rr|rrrrrrllllllll}
\hline
\hline
Case    &  $Sb$  & $\sigma_0 > 3$ & $\sigma_0 \le 3$ &  $N_i$& $p_i =x_i/N_i$ & $\tilde{x_1}$ & $\sigma_{x1}$ & $c^2$\phantom{1} & $\alpha$\\
\hline
Sample I & $\ge 4.5$  & $x_1 =\phantom{1} 5^a$   &  154 &  159 & 0.031 & 5.77 & 2.03 & 0.145 & 0.35\\
         & $<4.5$     & $x_2 = 17$  &  430 &  447 & 0.038 & \\
         & All        & $x_3 = 22$  &  584 &  606 & 0.036 & \\
Sample I & $\ge 5.0$  & 3   &  104 &  107 & 0.028 & 3.88 & 1.76 & 0.253 & 0.31\\
         & $<5.0$     & 19  &  480 &  499 & 0.038 & \\
         & All        & 22  &  584 &  606 & 0.036 & \\
Sample I & $\ge 6.0$  & 3   &   60 &   63 & 0.048 & 2.29 & 1.41 & 0.257 & 0.31\\
         & $<6.0$     & 19  &  524 &  543 & 0.035 & \\
         & All        & 22  &  584 &  606 & 0.036 & \\
\hline
         &            & $\sigma_0 >  0.57 Sb$ & $\sigma_0 \le 0.57 Sb$ \\ 
\hline
Sample I & $\ge 5.0$  & 2   &  105 &  107 & 0.019 & 6.36 & 2.22 & 3.85 & 0.025\\
         & $<5.0$     & 34  &  465 &  499 & 0.068 & \\
         & All        & 36  &  570 &  606 & 0.059 & \\
\hline
         &            & $\sigma_0 >  0.23 Sb + 0.2$ & $\sigma_0 \le 0.23 Sb+0.2$ \\ 
\hline
Sample II& $\ge 5.0$  & 1   &   28 &   29 & 0.034 & 3.43 & 1.64 & 2.189 & 0.069\\
         & $<5.0$     & 29  &  196 &  225 & 0.129 & \\
         & All        & 30  &  224 &  254 & 0.118 &                  \\
\hline
\end{tabular}

$^a$ Subscripts 1, 2, and 3 denote quantities for stars above
the dividing line, below it, and the total number, respectively. 

\end{table*}

To evaluate whether or not this absence
of binaries above $Sb = 5$~\kms\
has any statistical significance,
the $H_0$ hypothesis that the low $Sb$ and high $Sb$ regions  contain the
same fraction of stars  above the dashed line has been tested against the
alternate hypothesis that the frequencies  are different. The expected
number of binaries in the large $Sb$  region ($\tilde{x_1} = p_3 N_1$)
has been computed from the fraction
of binaries in the total sample (denoted $p_3$)
applied to the number of stars in the large $Sb$ region ($N_1$).
The corresponding standard deviation on $\tilde{x_1}$ (denoted $\sigma_{x1}$)
is computed from the expression for a
hypergeometric distribution, as applies to the variable $x_1$
(the number
of binaries in the large $Sb$ region), given the total number of stars in
the sample, $N_3$, the total number of binaries, $p_3 N_3$,
and the number of stars in the large $Sb$ region, $N_1$:
$ \sigma_{x1} = [N_1 p_3 (1 - p_3) (N_3 - N_1)/(N_3-1)]^{1/2}. $
The significance of the reduced, squared
difference $c^2 = [(\tilde{x_1} - x_1)/\sigma_{x1}]^2$ is computed from the
$\chi^2$ distribution with one degree of freedom. 
Table~\ref{Tab:hyper} lists the $c^2$ values and the corresponding values of the significance level 
$\alpha = 0.5\; {\rm Prob}_{c^2}^\infty (\chi^2)$ for different choices of the 
dividing line.
For sample~I, the stars flagged `Y' by
\citet{Famaey-2005} have been discarded and two versions of
the dividing line were considered: $\sigma_0 = 3$ \kms\ with
different $Sb$ thresholds and $\sigma_0 = 0.57 Sb$.
For sample II, the dividing line was $\sigma_0 = 0.23 Sb + 0.2$.
Table~\ref{Tab:hyper} reveals that the significance (denoted $\alpha$)
of the difference is at best about a few percent.

\begin{figure}
\includegraphics[width=\columnwidth]{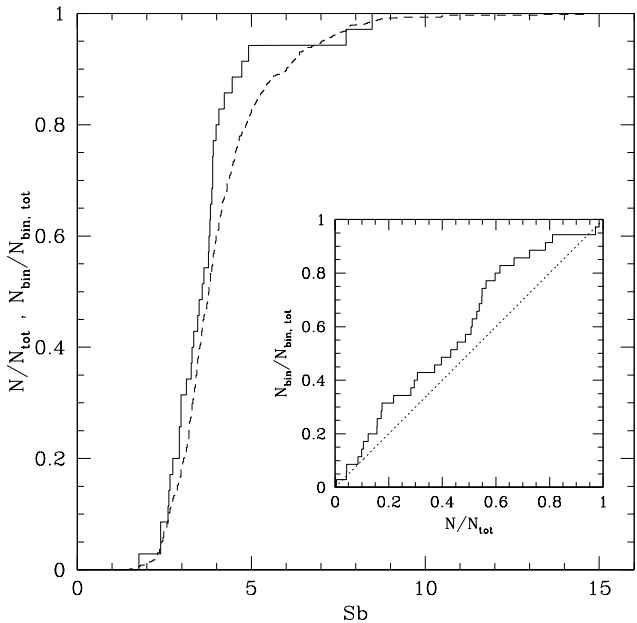}
\caption[]{\label{Fig:binfreq_KS}
The cumulative distributions of the total number of stars
and of the number of binaries, as a function of $Sb$, for sample~I
\protect\citep[after removing the visual binaries as well as stars flagged `Y'
by][]{Famaey-2005}.
The inset shows how the cumulative distribution of binaries varies throughout the
sample. If the frequency of binaries were the same for all values of $Sb$, the
solid line should follow the diagonal exactly.  
}
\end{figure}

A Kolmogorov-Smirnov test, based on the cumulative distributions displayed
in Fig.~\ref{Fig:binfreq_KS}, has been applied to the more populated
sample~I and confirms that significance level. With a total number of stars
of 606 (after removing the visual binaries and the `Y' group), there are 36
stars above the dividing line (having a slope of 0.57 in the
$Sb$ -- $\sigma_0(Vr)$ diagram). The maximum absolute difference between the
cumulative distributions of the total number of stars
and of the number of binaries, as a function of $Sb$, is 0.214
(Fig.~\ref{Fig:binfreq_KS}). This difference, clearly due to the lack of
`binaries' with $Sb \ge 5$~\kms, corresponds to a significance level of
7.5\% for rejecting the null hypothesis that the two distributions are
the same. 

\begin{figure}
\includegraphics[width=\columnwidth]{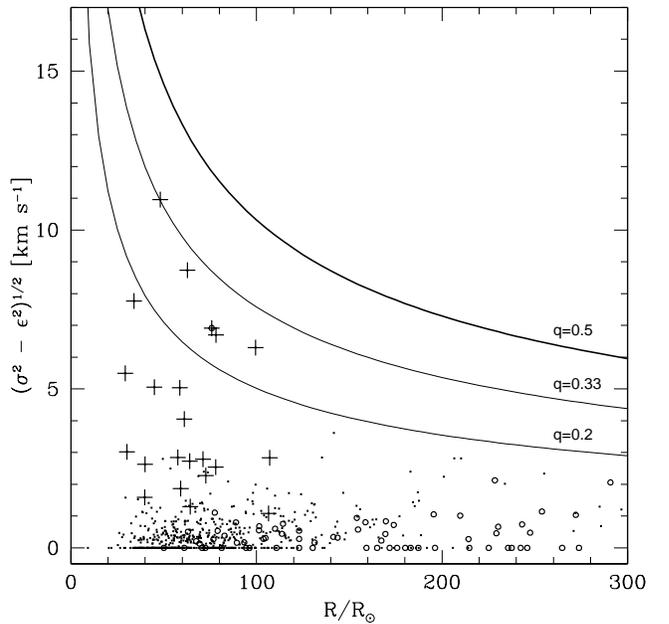}
\caption[]{\label{Fig:jit_vs_R}
The radial-velocity jitter versus stellar radius. Circles denote  
group `Y' stars and crosses binary stars with orbits
(binaries with no orbit,
i.e., stars flagged `0'
in \citet{Famaey-2005} are not included as they do not have maximum-likelihood
distances available, and hence radius is also not available). The curves correspond to the radial-velocity dispersion that
would be exhibited by a giant of 1.5~\Msun\ filling its Roche lobe in an
edge-on orbit (with a companion mass fixed by the mass ratio as labelled along
the curve). The theoretical radial-velocity semi-amplitude has been converted
into a standard dispersion by dividing by $\sqrt{2}$.
}
\end{figure}

The statistical significance of this lack of binaries is therefore not very
high. Yet, such a scarcity would be easy to understand.
First, as noted in Sect.~\ref{Sect:Sb}, $Sb$ is correlated with stellar
radius. At larger $R$, for given component masses, the minimum
possible orbital period (limited by RLOF) is longer and the maximum (edge-on)
amplitude of the orbital radial-velocity variations is {\em smaller}.
Any inclination would diminish the observed orbital motion even more, while
not affecting the intrinsic jitter.
Hence, a binary with a more extended giant would tend to hide below the
dividing line in the $Sb-\sigma_0$ diagram, among the intrinsic velocity
jitter. This point is illustrated by Fig.~\ref{Fig:jit_vs_R}.

Next, one should also notice that the RGB tip occurs around $Sb = 5$~\kms\ (see Fig.~\ref{Fig:HR_Sb}), thus
accounting for how the number of M giants drops dramatically for
$Sb \ge 5$~\kms, the RGB stars  living longer than asymptotic giant branch (AGB) stars. 
If any, a difference
in the binary properties of RGB and AGB stars could involve the large radii
reached on the upper RGB: (low-mass) AGB stars would thus be restricted to
long-period (i.e., small amplitude) systems, since the shorter-period
binaries would have gone through Roche-lobe overflow, with dramatic
consequences (dynamical mass transfer, common envelope, and orbital shrinkage
to end up as cataclysmic variables) when involving a deep convective envelope,
as is the case for RGB stars
(and with none of the physical processes to avoid orbital
shrinkage, advocated by
\citealt{Frankowski-2007a}
for RLOF involving AGB stars, being operative in RGB stars).

This interpretation is further supported by the period -- eccentricity diagram
($e$ -- $\log P$ diagram)
presented in Figs.~\ref{Fig:elogP-SB9} and \ref{Fig:elogP_cluster_mass},
revealing that no M giant is found at orbital periods shorter than 160~d. This limiting period 
corresponds to a Roche radius of 70~\Rsun\ when masses of 1.3~\Msun\
for the giant and 0.6~\Msun\ for the companion are adopted. 
In fact, no M giant binary lies to the left of the solid line in
Fig.~\ref{Fig:elogP-SB9}, representing the locus of a constant
periastron distance $A(1-e) = 157$~\Rsun\ which corresponds to
a Roche radius of 70~\Rsun\ {\em at periastron} for a system with
such masses (see Paper~III).
In contrast,
K giants, which are more compact, may be found at much shorter periods.

However, many M giants have radii that are much smaller than the 70~\Rsun\ threshold obtained above: Fig.~\ref{Fig:R_Sb} shows M giants
with radii as small as 30~\Rsun, in agreement with the spread observed by
\citet{VanBelle-1999} for the radii of M giants. Stars with such small radii could in principle be found to the left of the 70~\Rsun\
RLOF limit in the $e$ -- $\log P$ diagram, but
they are not. 
Part of the explanation could be that the observed envelopes in the
$e$--$\log P$ diagram are defined by tidal interactions more than by RLOF
-- and tidal interactions operate well before stars fill their Roche
lobe. This is especially indicated by the K giants, as they
contain a prominent circularised population around and below the
minimal period of their $e$--$\log P$ envelope
(Figs.~\ref{Fig:elogP-SB9},~\ref{Fig:elogP_cluster_mass}).
Among the M giant binaries, the short-period circular subpopulation is less
pronounced. Still, they stay within the 70~\Rsun\ envelope,
not extending down to Roche radii of $\sim 30$~\Rsun.

If the $e$--$\log P$ envelope were due to the tidal interactions,
should it not differ in shape from the periastron RLOF type envelope?
For comparison, Fig.~\ref{Fig:elogP-SB9} also displays short- and
long-dashed lines, corresponding to
the evolution of systems during circularisation, which leaves the angular
momentum per unit reduced mass [$h = A (1-e^2)$] constant
\citep{Zahn-1977,Hut81}.
Using Kepler's third law, this condition becomes $P^{2/3} (1-e^2)$ = constant
in the eccentricity -- period diagram.
The two dashed lines in Fig.~\ref{Fig:elogP-SB9} were plotted
adopting the same component masses of 1.3~\Msun\ and 0.6~\Msun, and with the
usual expression for the Roche radius \citep{Eggleton-83}, the Roche radii 
corresponding to the circular orbits at the base of the two lines in
Fig.~\ref{Fig:elogP-SB9} are 70 and 200~\Rsun.
However, lines of this shape do not follow the observed $e$--$\log P$
envelopes as closely as the lines of constant periastron distance do (especially 
in the upper panel of Fig.~\ref{Fig:elogP_cluster_mass}).
Either the shorter period objects are left outside the circularisation line
or a large empty area is included under this supposed envelope.
This may be explained as follows. Circularisation becomes fast when the
giant comes close to filling its Roche lobe at periastron. But since   
the circularisation path in
the $e$ -- $\log P$ diagram is steeper than the periastron line
(see Fig.~\ref{Fig:elogP-SB9}), the system evolves somewhat away
from the periastron line and thus away from the strong
circularisation regime. It does not ``slide'' all the way down along the
circularisation path, because it loses the driver. Only when other effects
bring the star and its Roche lobe closer again does circularisation also
accelerate again. It should only be fast close to the the periastron
envelope.
Of these two lines, it is thus the periastron envelope that would regulate
the evolution of the $e$ -- $\log P$ diagram of an ensemble of systems.   

In any case, the minimal Roche lobe radii inferred from the M giant
$e$ -- $\log P$ diagram are apparently much too large for the measured
radii for M giants.
They correspond to Roche lobe-filling factors $R/R_R$, which are much
too small for circularisation to operate efficiently (i.e., on a time scale
comparable to the RGB evolutionary time scale). Interestingly enough, a similar
difficulty has been pointed out by 
\citet{Muerset-99} and \citet{Mikolajewska-2007} in the context of symbiotic stars. 
The former authors have
noticed that, surprisingly, the M components in symbiotic stars rarely had
Roche lobe-filling factors over 0.6, whereas the latter author
notes that symbiotic stars exhibit ellipsoidal variability despite their
moderate Roche lobe-filling factors. A possible explanation for this discrepancy
could be that the usual
expression for the Roche radius \citep{Eggleton-83} overestimates it
due to neglecting radiation pressure and other factors countering   
stellar gravity,  which can both shrink the effective Roche-lobe and make
surface ellipsoidal distortions important at smaller radii
\citep{Schuerman-1972, Frankowski-2001}. This
possibility is explored further by \citet{Dermine-Jorissen-2008}.
The observed limit on the giant's Roche-lobe would be explained by a
reduction of the effective gravity on the stellar surface to 0.15-0.35 of
its Newtonian value \citep{Frankowski-2001,Dermine-Jorissen-2008}.

From the above discussion one may conclude that 
the frequency of binaries must be expected to be lower among M giants than among K giants, because the larger radii of the former forbid them from being located to the left of the dashed line in the
$e$ -- $\log P$ diagram of Fig.~\ref{Fig:elogP-SB9}, whereas this restriction does not apply to K giants. This statement will be quantified in Sect.~\ref{Sect:K}.

\subsection{Comparison with K giants}
\label{Sect:K}

\begin{figure}
\includegraphics[width=\columnwidth]{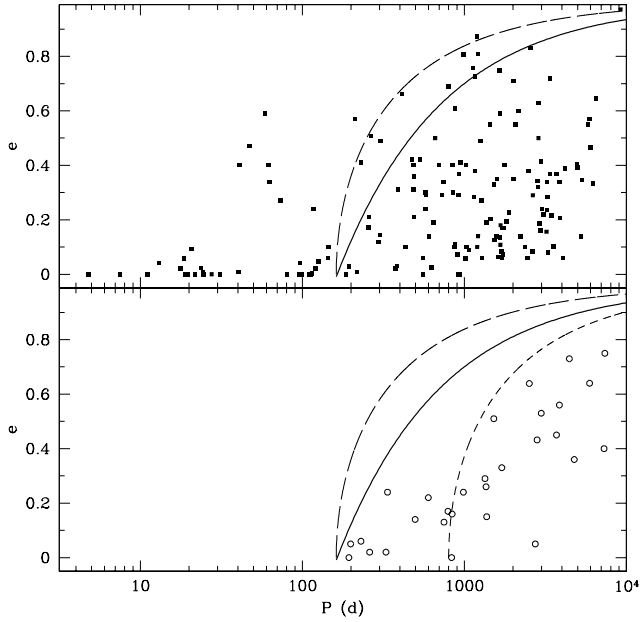}
\caption[]{\label{Fig:elogP-SB9}
Top panel: The distribution in the eccentricity -- period diagram of the 164
binaries with a KIII primary extracted from the Ninth Catalogue of Spectroscopic
Binary Orbits \citep{Pourbaix-04a}. The dashed lines correspond to
the evolution of the systems during circularisation. The thick solid
line represents the locus of constant periastron
distance $A(1-e)=157$~\Rsun, corresponding to a Roche radius at periastron of 70~R$_\odot$, assuming
masses of 1.3 and 0.6~\Msun\ for the giant and its companion, respectively.
Most of the binaries with $P < 100$~d  are fast-rotating, active RS CVn stars.
Some of these systems are not yet tidally circularised. They could actually be {\it pre-main-sequence stars} (PMS), since RS~CVn and PMS stars are sometimes confused \citep[e.g.,][]{Rucinski-1977}.
Star 28  from \citet{Torres-2002} is one such case ($P = 40.9$~d, $e = 0.40$, spectral type K3III).  
Bottom panel: Same as top for the M giants. The short-dashed line corresponds to evolution of longer-period 
systems during circularisation (see Sect.~\ref{Sect:above5} for 
discussion of the lines). 
}

\end{figure}

 \begin{figure}
 \includegraphics[width=\columnwidth]{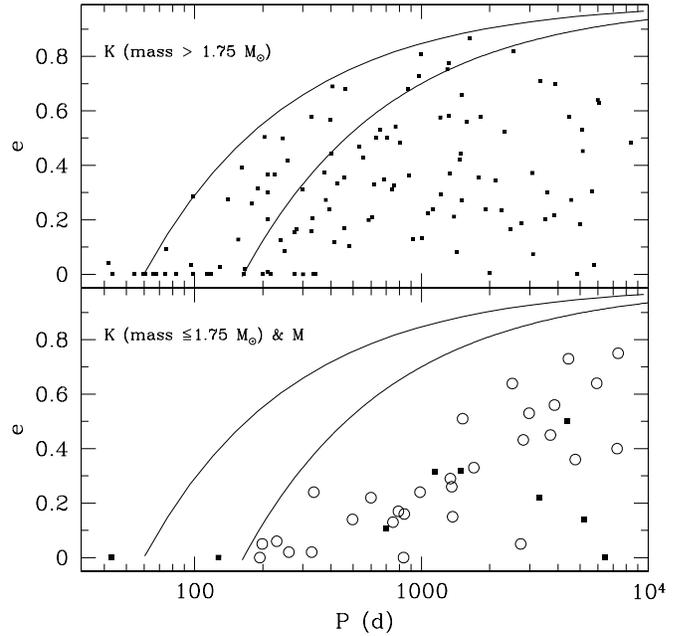}
 \caption[]{\label{Fig:elogP_cluster_mass}
 Top panel: The distribution in the eccentricity -- period diagram of K
 giants  in clusters with a turnoff mass higher than 1.75~\Msun. Data from \citet{Mermilliod-2007b}. The solid lines represent the loci of constant periastron
distances $A(1-e)=79$ and 157~\Rsun, corresponding to a Roche radius at periastron of 35 and 70~R$_\odot$, respectively.
 Bottom panel: Same as top for the M giants (open circles) and K giants (filled squares) in clusters with
 a turnoff mass lower than or equal to  1.75~\Msun.
 }
 \end{figure}

\begin{table*}
\caption[]{\label{Tab:frequency}
Summary of the binary frequency among different samples of K and M giants.
}
\begin{tabular}{lllllll}
\hline
\hline
Stellar type & Sample properties & $N_{\rm tot}$ & binary frequency (\%) & Ref.\\
\hline\\
Field MIII & I (whatever $N\ge 2$, all $Sb$) & 771 & 5.6 & Sect.~\ref{Sect:freqI}\\
           & II, III, IV ($N > 4$, all $Sb$) & 254 & 9.8 (+2.4?) & Sect.~\ref{Sect:freqII}\\
Field MIII (without Miras)  & I (whatever $N\ge2$, $Sb < 5$) & 603 & 6.3 & Sect.~\ref{Sect:freqI}\\
           & II, III, IV ($N > 4, Sb < 5$)  & 225 & 11.1 (+2.7?) & Sect.~\ref{Sect:freqII}\\
Field KIII & (whatever $N\ge2$) & 5643$^a$ & 14.5$^b$& \citet{Famaey-2005}\\
           & Mermilliod \& Mayor (2007) ($N = 3$) & 285 & $16\pm2$ \\ 
Cluster KIII & Mermilliod \& Mayor (2007) ($N = 3$) & 704 & $30.8\pm1.7$ \\
\hline
\end{tabular}

$^a$ This number differs from the one quoted by \citet{Famaey-2005}, because we
restrict the sample to stars having at least two radial-velocity measurements.\\
$^b$ at the level Prob$(\chi^2(Vr)) \le 0.001$
\end{table*}

Not many estimates of the binary frequency among K giants exist in the literature. An
early study by \citet{Harris-McClure-83} of 40 K giants in the field resulted in
a frequency of 15 to 20\% spectroscopic binaries (over a 3~yr time
span and with a radial-velocity internal error of 0.40~\kms).  In open
clusters, Mermilliod \& Mayor (2008, in prep.) find a much higher
frequency of $30.8\pm1.7$\% (= 217/704) for G and K giants under similar observing conditions
(similar time span with 3 measurements per star). Surprisingly, for the giants that are not cluster members, the
same authors derive a much lower binary frequency, namely $16\pm2$\% (= 46/285).

A binary frequency may also be derived for the sample of Hipparcos K giants monitored with CORAVEL and studied by \citet{Famaey-2005}.
Table~\ref{Tab:frequency} lists the binary frequency  among that sample of K giants. To compare binary frequencies among K and M giants, one should try to avoid  systematic and selection effects as much as possible.  Most importantly, the comparison should involve binary frequencies derived from the same number of measurements covering approximately the same time span. For M giants, Table~\ref{Tab:frequency} reveals that not respecting this condition may cause the binary frequency estimate to vary by
about a factor of two
(from 6.3\% in sample~I to 11.1\% in samples II, III, IV, after excluding the Miras).
Therefore, the binary frequencies among sample I of M giants and among K giants were re-assessed based on the radial-velocity standard deviations computed only from 2 datapoints (the first and the last available). 
The stars flagged as binaries in a given sample may then change. However, it is found that the {\em total number} of binaries remains the same (to within 2 units), both for K and M giants.
Therefore, we may conclude that the fraction of spectroscopic
binaries among M giants (after correcting for the difficulty of finding
binaries among Mira variables) is less than among K giants by a factor 6.3/14.5= 0.43.

What could  be the origin of such a difference in the binary frequencies among
K and M giants? 
Part of it probably comes from the greater difficulty of finding 
binaries among M giants because of their smaller orbital velocity
amplitudes and larger intrinsic jitter (Fig.~\ref{Fig:jit_vs_R}), which
prevents a small
velocity difference from being ascribed to the orbital motion as it is for K
giants. This is especially true for a small number of measurements, as in our
sample~I.

Another possibility may reside in the different distributions of K and M giants in the eccentricity -- period diagram, as shown in Fig.~\ref{Fig:elogP-SB9}. 
It must be stressed that this figure refers to samples of {\em field} giants, i.e., orbits for K giants being retrieved from the Ninth Catalogue of Spectroscopic Binary
Orbits \citep{Pourbaix-04a}, explicitely excluding the orbits for K giants in clusters derived
by \citet{Mermilliod-2007b} (Fig.~\ref{Fig:elogP_cluster_mass}). This allows for a direct comparison of this sample with the sample of M giants without worrying about differences between field and cluster populations.

For stars evolving on the first-ascent giant branch,
the radii of K giants are on average smaller than those of M giants, so that the
minimum orbital period observed for M giants is expected to be somewhat longer than for
K giants, as confirmed by Fig.~\ref{Fig:elogP-SB9}. The resulting decrease in
binary frequency may be estimated by computing the ratio between the number
 of K giants to the right of the solid line to the total number
of K giants, or 118/164 = 0.72. Applying this factor to the frequency
of M giant binaries thus yields 6.3 / 0.72 = 8.8~\%, closer to 
the binary frequency among field K giants, but still well below it,
and this result is robust. For example, if the long-dashed line
from Fig.~\ref{Fig:elogP-SB9} is used as the limit,
the correction factor would be 126/164 = 0.77 instead of 0.72.

Still another  bias,  which may alter the factor 0.72 used above, should be
taken into consideration. This bias is related to the evolutionary status of
the  K giant, which may either be the first-ascent giant branch or the core-He  burning. For low-mass stars, the latter phase implies that stars
have gone  through the tip of the RGB where they reached a very large radius
(similar to  that of M giants). Therefore, core-He-burning K giants
are expected to  have an eccentricity -- period distribution similar to
that of M giants.

Thus the eccentricity -- period diagram of {\em low-mass} K giants must be
expected to mix such K giants with orbital properties similar to M giants
with K giants having orbital parameters  more typical of less evolved
(first-ascent giant branch) stars  (i.e., short periods and possibly high
eccentricities). Since stars spend more time in the He-burning clump, the former
should be more numerous, however. For intermediate-mass  K giants, this
segregation does not occur since they never went through a stage  with
large radii. 
This may be checked by using the sample  of K giant binaries in open
clusters \citep[][ excluding stars flagged as
non-members]{Mermilliod-2007b}, where the  K-giant mass may be identified
with the cluster turnoff mass.
Figure~\ref{Fig:elogP_cluster_mass} reveals that, as expected, the K giants
in clusters with  turnoff masses lower than 1.75~\Msun\ are mostly found
in the region occupied by M giants, with only two short-period circular
orbits (out of 9), in accordance with the masses of M giants estimated to
fall in the range  1.2 -- 1.7 \Msun\ from Fig.~\ref{Fig:HR_Sb}.
The situation looks quite different indeed when considering
intermediate-mass K giants (upper panel of
Fig.~\ref{Fig:elogP_cluster_mass}). If the sample of K giant binaries were
only made out of low-mass stars, the correcting factor would then be
7/9 = 0.78 instead of 83/126 = 0.66
for the intermediate-mass K giants
(close to the value 0.72 derived above for the total sample of {\it field} K giants). 
The true correction factor should thus be the average of these two values, weighted by the (unknown) 
fraction of low-mass with respect to intermediate-mass stars among
the sample of field K-giant binaries. This true correction
factor should be in the range 0.6--0.8 (irrespective of the exact weighing function), 
which is still too much to
account for the factor of 2--3 difference between the binary frequencies
among K and M giants. In the end, it is the difficulty of finding
spectroscopic binaries among M giants due to their larger radial-velocity
jitter that must be
invoked to account for the  binaries missing among M giants.

\section{Conclusions}

In this paper we have derived the frequency of spectroscopic binaries among
field M giants, for the first time based on an extensive sample, 771 M
giants in total. The frequency obtained, 6.3\%, 
is much lower than for K
giants (ranging from 14 to 31\%, depending on the samples
considered). 
However, the binary frequency derived from this large sample of M giants is not very meaningful, since there are important observational difficulties with detecting M giant binaries, because on average they
have smaller orbital velocity amplitudes (longer periods) and
larger intrinsic velocity jitter than K giants.
This effect is especially important for samples with only a fewmeasurements per star, as 
is the case with this sample of 771 M giants.

A higher binary frequency was obtained in a smaller M giant sample with more
radial-velocity measurements per object: 11.1\%  confirmed plus 2.7\% possible
binaries. Even though this frequency is close to the lower bound of the binary frequency among K giants, we have shown that they cannot be directly compared because they were obtained under different observing conditions (mostly the number of observations per object). When one tries to compare the binary frequencies for samples of K and M giants under similar observing conditions, the binary frequency among M giants remains lower than that among K giants by a factor of about 2. Part of this difference may stem from the destruction of shorter-period K giant binaries due to catastrophic RLOF leading to common envelope,
before they can become M giant binaries. However, how important this effect is
depends on the mass distribution in the considered population.

The lack of M-giant  binaries with orbital periods shorter than 160~d, as compared to K-giant binaries that have orbital periods as short as 4~d, has been clearly demonstrated from our extensive set of orbital elements. 

We have found that the CORAVEL line-width parameter $Sb$ is better
correlated with the stellar radius than with either luminosity or effective
temperature separately. This allowed identification of the $R$ -- $Sb$ relation
outliers HD~190658 and HD~219654 as fast rotators, possibly due to the binary
companion influence; indeed, HD~190658 and HD~219654 turn out to be known as
spectroscopic binaries. A third candidate was rejected, because its large $Sb$
was caused by light contamination from a close visual companion.

Finally, we have shown that M giants are circularised even though their
Roche lobe-filling factor, estimated from the usual expression for the Roche
radius, is about 0.5, in contrast to predictions from tidal
circularisation theory. A similar difficulty has been reported for
ellipsoidal variables among symbiotic systems. We argue that these
discrepancies call for a revision of the Roche-lobe concept in the presence
of radiation pressure from the giant component.

\acknowledgements{
Stimulating discussions with H. Van Winckel helped to improve this paper.
This work was partly funded by an {\it Action de recherche concert\'ee (ARC)} from the {\it Direction g\'en\'erale de l'Enseignement non obligatoire et de la Recherche scientifique -- Direction de la recherche scientifique -- Communaut\'e fran\c caise de Belgique.}
}
\bibliographystyle{apj} 
\bibliography{ajorisse_articles} 

\begin{thebibliography}{}

\bibitem[\protect\astroncite{{Alvarez} et~al.}{2001}]{Alvarez-2001}
{Alvarez} R., {Jorissen} A., {Plez} B., {Gillet} D., {Fokin} A., {Dedecker} M.
  2001, \aap, 379, 305

\bibitem[\protect\astroncite{{Benz} \& {Mayor}}{1981}]{Benz-1981}
{Benz} W., {Mayor} M. 1981, \aap, 93, 235

\bibitem[\protect\astroncite{{Bessell} et~al.}{1998}]{Bessell-98}
{Bessell} M.~S., {Castelli} F., {Plez} B. 1998, \aap, 333, 231

\bibitem[\protect\astroncite{{Charbonnel} et~al.}{1996}]{Charbonnel-1996}
{Charbonnel} C., {Meynet} G., {Maeder} A., {Schaerer} D. 1996, A\&AS, 115, 339

\bibitem[\protect\astroncite{{Cowley}}{1969}]{Cowley-1969}
{Cowley} A.~P. 1969, \pasp, 81, 297

\bibitem[\protect\astroncite{{De Medeiros} et~al.}{1996}]{DeMedeiros-1996}
{De Medeiros} J.~R., {Da Rocha} C., {Mayor} M. 1996, \aap, 314, 499

\bibitem[\protect\astroncite{{De Medeiros} et~al.}{2002}]{DeMedeiros-2002}
{De Medeiros} J.~R., {Da Silva} J.~R.~P., {Maia} M.~R.~G. 2002, \apj, 578, 943

\bibitem[\protect\astroncite{{De Medeiros} \& {Mayor}}{1999}]{DeMedeiros-1999}
{De Medeiros} J.~R., {Mayor} M. 1999, \aaps, 139, 433

\bibitem[\protect\astroncite{{Dermine} et~al.}{2009}]{Dermine-Jorissen-2008}
{Dermine} T., {Jorissen} A., {Siess} L., {Frankowski} A. 2009, \aap, submitted

\bibitem[\protect\astroncite{{Eggleton}}{1983}]{Eggleton-83}
{Eggleton} P.~P. 1983, \apj, 268, 368

\bibitem[\protect\astroncite{{Famaey} et~al.}{2005}]{Famaey-2005}
{Famaey} B., {Jorissen} A., {Luri} X., {Mayor} M., {Udry} S., {Dejonghe} H.,
  {Turon} C. 2005, \aap, 430, 165

\bibitem[\protect\astroncite{{Famaey} et~al.}{2009}]{Famaey-2008:b}
{Famaey} B., {Pourbaix} D., {Jorissen} A., {Frankowski} M., {Van Eck} S.,
  {Mayor} M., {Udry} S. 2009, \aap, this volume (Paper I)

\bibitem[\protect\astroncite{{Frankowski} et~al.}{2007}]{Frankowski-2007}
{Frankowski} A., {Jancart} S., {Jorissen} A. 2007, \aap, 464, 377

\bibitem[\protect\astroncite{{Frankowski} \&
  {Jorissen}}{2007}]{Frankowski-2007a}
{Frankowski} A., {Jorissen} A. 2007, Baltic Astronomy, 16, 104

\bibitem[\protect\astroncite{{Frankowski} \& {Tylenda}}{2001}]{Frankowski-2001}
{Frankowski} A., {Tylenda} R. 2001, \aap, 367, 513

\bibitem[\protect\astroncite{{Harris} \& {McClure}}{1983}]{Harris-McClure-83}
{Harris} H.~C., {McClure} R.~D. 1983, ApJ, 265, L77

\bibitem[\protect\astroncite{{Hut}}{1981}]{Hut81}
{Hut} P. 1981, \aap, 99, 126

\bibitem[\protect\astroncite{{Jorissen}}{2003}]{Jorissen-03}
{Jorissen} A. 2003,
\newblock in H. {Habing}, H. {Olofsson} (eds.), Asymptotic Giant Branch Stars,
  Springer Verlag, New York, p.~461

\bibitem[\protect\astroncite{{Jorissen} et~al.}{2009}]{Jorissen-2008}
{Jorissen} A., {Frankowski} A., {Famaey} B., {Van Eck} S. 2009, \aap, this
  volume (Paper III)

\bibitem[\protect\astroncite{{Jorissen} et~al.}{1998}]{Jorissen-VE-98}
{Jorissen} A., {Van Eck} S., {Mayor} M., {Udry} S. 1998, \aap, 332, 877

\bibitem[\protect\astroncite{{Leedjaerv}}{1998}]{Leedjaerv-1998}
{Leedjaerv} L. 1998, \aap, 338, 139

\bibitem[\protect\astroncite{{Lucke} \& {Mayor}}{1982}]{Lucke-Mayor-1982}
{Lucke} P.~B., {Mayor} M. 1982, \aap, 105, 318

\bibitem[\protect\astroncite{{Mermilliod} et~al.}{2007}]{Mermilliod-2007b}
{Mermilliod} J.-C., {Andersen} J., {Latham} D.~W., {Mayor} M. 2007, \aap, 473,
  829

\bibitem[\protect\astroncite{{Miko\l ajewska}}{2007}]{Mikolajewska-2007}
{Miko\l ajewska} J. 2007, Baltic Astron., 16, 1

\bibitem[\protect\astroncite{{M{\"u}rset} \& {Schmid}}{1999}]{Muerset-99}
{M{\"u}rset} U., {Schmid} H.~M. 1999, A\&AS, 137, 473

\bibitem[\protect\astroncite{{Platais} et~al.}{2003}]{Platais-2003}
{Platais} I., {Pourbaix} D., {Jorissen} A., {Makarov} V.~V., {Berdnikov} L.~N.,
  {Samus} N.~N., {Lloyd Evans} T., {Lebzelter} T., {Sperauskas} J. 2003, \aap,
  397, 997

\bibitem[\protect\astroncite{{Pourbaix} et~al.}{2004}]{Pourbaix-04a}
{Pourbaix} D., {Tokovinin} A.~A., {Batten} A.~H., {Fekel} F.~C., {Hartkopf}
  W.~I., {Levato} H., {Morrell} N.~I., {Torres} G., {Udry} S. 2004, \aap, 424,
  727

\bibitem[\protect\astroncite{{Rucinski}}{1977}]{Rucinski-1977}
{Rucinski} S.~M. 1977, \pasp, 89, 280

\bibitem[\protect\astroncite{{Samus}}{1997}]{Samus-1997}
{Samus} N.~N. 1997, Informational Bulletin on Variable Stars, 4501, 1

\bibitem[\protect\astroncite{{Schaller} et~al.}{1992}]{Schaller-1992}
{Schaller} G., {Schaerer} D., {Meynet} G., {Maeder} A. 1992, A\&AS, 96, 269

\bibitem[\protect\astroncite{{Schuerman}}{1972}]{Schuerman-1972}
{Schuerman} D.~W. 1972, Ap\&SS, 19, 351

\bibitem[\protect\astroncite{{Torres} et~al.}{2002}]{Torres-2002}
{Torres} G., {Neuh{\"a}user} R., {Guenther} E.~W. 2002, \aj, 123, 1701

\bibitem[\protect\astroncite{{Udry} et~al.}{1998a}]{Udry-98a}
{Udry} S., {Jorissen} A., {Mayor} M., {Van Eck} S. 1998a, A\&AS, 131, 25

\bibitem[\protect\astroncite{{Udry} et~al.}{1997}]{Udry-1997}
{Udry} S., {Mayor} M., {Andersen} J., {Crifo} F., {Grenon} M., {Imbert} M.,
  {Lindegren} H., {Maurice} E., {Nordstroem} B., {Pernier} B., {Prevot} L.,
  {Traversa} G., {Turon} C. 1997,
\newblock in M. {Perryman} (ed.), Hipparcos - Venice '97 (ESA SP-402), ESA,
  Noordwijk, p.~693

\bibitem[\protect\astroncite{{Udry} et~al.}{1998b}]{Udry-98b}
{Udry} S., {Mayor} M., {Van Eck} S., {Jorissen} A., {Prevot} L., {Grenier} S.,
  {Lindgren} H. 1998b, A\&AS, 131, 43

\bibitem[\protect\astroncite{{van Belle} et~al.}{1999}]{VanBelle-1999}
{van Belle} G.~T., {Lane} B.~F., {Thompson} R.~R., {Boden} A.~F., {Colavita}
  M.~M., {Dumont} P.~J., {Mobley} D.~W., {Palmer} D., {Shao} M., {Vasisht}
  G.~X., {Wallace} J.~K., {Creech-Eakman} M.~J., {Koresko} C.~D., {Kulkarni}
  S.~R., {Pan} X.~P., {Gubler} J. 1999, \aj, 117, 521

\bibitem[\protect\astroncite{{Van Eck} \&
  {Jorissen}}{2000}]{VanEck-Jorissen-00}
{Van Eck} S., {Jorissen} A. 2000, \aap, 360, 196

\bibitem[\protect\astroncite{{Wielen} et~al.}{1999}]{Wielen-1999}
{Wielen} R., {Dettbarn} C., {Jahrei\ss} H., {Lenhardt} H., {Schwan} H. 1999,
  A\&A, 346, 675

\bibitem[\protect\astroncite{{Wood}}{1990}]{Wood-1990}
{Wood} P.~R. 1990,
\newblock in M.-O. {Mennessier}, A. {Omont} (eds.), From Miras to Planetary
  Nebulae: Which Path for Stellar Evolution?, Editions Fronti\`eres,
  Gif-sur-Yvette, ~67

\bibitem[\protect\astroncite{{Zahn}}{1977}]{Zahn-1977}
{Zahn} J.-P. 1977, \aap, 57, 383

\end{thebibliography}

\end{document}